\documentclass[]{jfm}
\graphicspath{{./}}

\newcommand{\aver}[1]{ \! \left\langle {#1} \right \rangle \!}
\newcommand{\DR}{\mathcal R }

\lefttitle{M.Quadrio \etal}
\righttitle{Journal of Fluid Mechanics}

\begin{document}

\title{On the optimal period of spanwise wall forcing \\ for turbulent drag reduction}

\author{Maurizio Quadrio, Federica Gattere, Marco Castelletti \and Alessandro Chiarini}
\affiliation{Dipartimento di Scienze e Tecnologie Aerospaziali, Politecnico di Milano, via La Masa 34, 20156 Milano, Italy}

\maketitle

\begin{abstract}
Turbulent channel flow controlled by spanwise wall oscillations is studied using direct numerical simulations to improve how spanwise forcing reduces skin-friction drag. Harmonic wall oscillations generate a periodic transverse Stokes layer whose thickness $\delta$ is determined by the forcing period $T$. Although an optimal $T$ that maximizes drag reduction is known to exist, its physical significance remains unclear.
To elucidate it, we extend the spanwise Stokes layer by augmenting wall oscillation with an additional spanwise body force. In this formulation, $\delta$ and $T$ become decoupled and can be varied independently. The oscillating wall thus appears as a special and suboptimal case of spanwise forcing. Optimal performance is obtained for substantially smaller $T$ and larger $\delta$ than those of the classical Stokes layer. For the conditions examined, with Reynolds number and forcing amplitude held fixed, the maximum drag reduction increases by approximately one third, while the maximum net energy saving improves markedly from $-35\%$ to $+16\%$.
These findings suggest that drag-reduction strategies based on spanwise forcing deserve renewed scrutiny: wall oscillation represents only one possible actuation method, and not necessarily the most effective one.
\end{abstract}


\section{Introduction}
\label{sec:intro}

Reducing turbulent skin-friction drag is a central objective in fluid mechanics, because of environmental and economic considerations. Over the years, several active and passive flow control strategies have been considered. Among these, wall-based techniques that operate without feedback, relying solely on predetermined actuation, offer an attractive balance between simplicity of implementation and potential for energy saving. 

In this work, we focus on spanwise wall forcing \citep[see][for a review]{ricco-skote-leschziner-2021}, which has been proved to remain effective at high Reynolds numbers \citep{gatti-etal-2025} and Mach numbers \citep{gattere-etal-2024}, and to provide large net savings. 
The simplest and earliest variant of spanwise forcing is the spatially uniform spanwise oscillation of a plane wall \citep{jung-mangiavacchi-akhavan-1992}. Other, more efficient variants share the same underlying working principle.
The wall oscillates harmonically in the spanwise direction as a function of time $t$ as
\begin{equation}
  w_w(t) = A \sin \left( \frac{2 \pi}{T} t \right),
\label{eq:OW}
\end{equation}
where $w_w$ is the spanwise velocity component of the wall (the other components are set to zero), and $A$ and $T$ denote the amplitude and period of the oscillation, respectively. 
The oscillation \eqref{eq:OW} generates a periodic spanwise cross-flow which interacts with the turbulent flow, eventually reducing skin-friction drag. \cite{quadrio-sibilla-2000} were first to realize that the phase-averaged spanwise turbulent velocity profile follows the laminar solution $w_{\text{\tiny \!\! SL}}(y,t)$ of the second Stokes problem, hereafter referred to as the Stokes layer (SL), with small deviations only occurring for large $T$. 
It is a textbook result that an indefinite plane wall oscillating harmonically as in \eqref{eq:OW} beneath a still fluid generates a time-varying velocity profile given by
\begin{equation}
  w_{\text{\tiny \!\! SL}}(y,t) = 
  A \exp \left( - \frac{y}{\delta} \right) 
  \sin \left( \frac{2\pi}{T} t - \frac{y}{\delta} \right),
\label{eq:SL}
\end{equation}
where $y$ is the wall-normal coordinate and $\delta$ is the SL thickness, conventionally defined as the wall-normal distance at which the maximum spanwise velocity during the cycle decays to $\exp(-1)$ times the wall value $A$.
Owing to the linearity of the equations governing the Stokes problem, the amplitude $A$ simply acts as a scaling factor; the shape of the SL is described by the parameters $T$ and $\delta$. However, they are not independent, since $\delta$ is determined by the period $T$ (and by the fluid kinematic viscosity, that will not be explicitly considered as a control parameter hereinafter) as
\begin{equation}
\delta = \delta_{\text{\tiny \!\! SL}}(T; \nu) \equiv \sqrt{\frac{\nu T}{\pi}}.
\label{eq:deltaSL}
\end{equation}

Although the SL has been extensively studied, its interaction with turbulence remains not fully understood.
Since the earliest numerical studies on spanwise forcing, the existence of an optimal oscillation period $T_{opt}$ that maximizes drag reduction has been empirically established.
Since the amount of drag reduction is proportional to the forcing amplitude, $T_{opt}$ is often defined under the condition of constant $A$ and denoted $T_{opt,A}$. An alternative definition keeps the peak wall displacement $D = A T / \pi$ constant, leading to $T_{opt,D}$ \citep{quadrio-ricco-2004}. The latter choice is particularly relevant in experimental setups, where hardware may impose physical constraints on the maximum displacement \citep{laadhari-skandaji-morel-1994, choi-2002, gatti-etal-2015, marusic-etal-2021}. 
However, most studies have focused on $T_{opt,A}$, which for simplicity will be referred to in this paper by omitting the subscript $A$.

There is broad consensus that $T_{opt}$ scales in viscous units, with a value of $T_{opt}^+ \approx 100$, where the superscript $^+$ denotes viscous scaling based on the fluid viscosity and the friction velocity of the uncontrolled flow. Since skin-friction drag reduction directly alters the friction level, a viscous scaling based on the actual value of the friction velocity in the controlled flow is meaningful \citep{quadrio-2011}. Actual viscous scaling is indicated by the superscript $^*$, and leads to $T_{opt}^* \approx 75$. According to \eqref{eq:deltaSL}, the value of $T_{opt}$ implies an optimal Stokes layer thickness of $\delta_{opt}^+ \approx 5.6$ or $\delta_{opt}^* \approx 4.9$. 
The value of $T_{opt}$ should be interpreted with caution, as its determination is delicate. The optimum lies on a relatively flat region of the drag-reduction curve, so even minor measurement inaccuracies can shift the estimated value of $T_{opt}$ significantly. Additionally, in numerical simulations the choice of scaling is intertwined with the type of comparison used between the controlled and uncontrolled flows: either the flow rate (CFR) or the pressure gradient (CPG) can be kept constant \citep{quadrio-frohnapfel-hasegawa-2016}, leading to non-negligible differences, especially at low Reynolds numbers. 
However, the currently accepted values of $T_{opt}$ are quite robust. 
For example, \citet{choi-xu-sung-2002} employed direct numerical simulations (DNS) of turbulent channel flow to show that the optimal oscillation period is $T_{opt}^+ \approx 100$ across a range of forcing amplitudes and friction Reynolds numbers $Re_\tau$.
For a turbulent channel flow at $Re_\tau = 200$, \citet{quadrio-ricco-2004} confirmed via DNS that, for a given $A^+$, the maximum drag reduction is achieved when $T^+ = 100 - 125$.
\cite{touber-leschziner-2012} and \cite{agostini-touber-leschziner-2014} found $T^+_{opt} \approx 100$ at the larger Reynolds numbers of $Re_\tau = 500$ and $Re_\tau = 1000$. 
\citet{gatti-quadrio-2016} confirmed the optimal value through a comprehensive DNS study spanning several forcing amplitudes and Reynolds numbers up to $Re_\tau = 1000$; the analysis has been recently extended up to $Re_\tau = 6000$ \citep{gatti-etal-2025}.
\cite{gattere-etal-2024} showed that the optimal value $T^+ \approx 100$ holds also in the compressible regime, at least up to a Mach number of $1.5$.
Several experimental works, although typically affected by the maximum-displacement limitation discussed above, have indirectly confirmed the value of $T_{opt}^+$ over a range of Reynolds numbers and forcing amplitudes: examples are \cite{laadhari-skandaji-morel-1994, trujillo-bogard-ball-1997, gatti-etal-2015, kempaiah-etal-2020}. 

Despite this extensive body of work, the physical significance of the optimal values $T_{opt}^+ \approx 100$ and $\delta_{opt}^+ \approx 5.7$ remains unsettled.
In fact, the time scale $T_{opt}$ can be linked to several characteristic flow scales. For example, the typical Lagrangian timescale expressing the lifetime of the near-wall coherent structures measured in the buffer layer is 75 viscous time units \citep{quadrio-luchini-2003}. 
Also, the bursting phase of the near-wall cycle described by \cite{jimenez-2013} is about $100$ viscous units.
Moreover, since near-wall turbulence in the buffer layer convects at approximately $10$ times the friction velocity \citep{kim-hussain-1993}, $T_{opt}$ defines a longitudinal length scale $L_{opt}^+ \approx 1000$. In the oscillating-wall study by \citet{touber-leschziner-2012}, this has been associated with the characteristic length of near-wall low-velocity streaks.
$T_{\rm opt}$ can also be interpreted through the maximum spanwise displacement of the oscillating wall, $D^+$, that matches the mean spanwise spacing of the streaks \citep{choi-xu-sung-2002}.
Alternatively, $T_{opt}$ can be converted into a wall-normal velocity scale, namely the thickness $\delta_{opt}$ of the Stokes layer, through equation \eqref{eq:deltaSL}. The optimal thickness may be interpreted as a diffusion length scale, indicating how far the Stokes layer penetrates into the bulk flow, or as a length setting the mean spanwise wall shear. It is worth noting that the concept of penetration length combines the thickness \eqref{eq:deltaSL} with the actual level of spanwise velocity that survives at $y=\delta$. In previous studies, the optimal Stokes layer thickness has been linked to the effectiveness of the oscillating wall, supporting the qualitative concept \citep{baron-quadrio-1996, ricco-2004} that spanwise oscillations weaken the connection between low-speed streaks (residing at $y^+ \lesssim 10$) and quasi-streamwise vortices, which exist further from the wall.
These competing interpretations of $T_{opt}$, albeit all referring to near-wall turbulence, together with our inability to discriminate among them, hint at a limited understanding of the overall drag reduction mechanism induced by the oscillating wall.

The aim of this work is to separately describe the role of $T$ and $\delta$ in the reduction of drag. The constraint $\delta = \delta_{\text{\tiny \!\! SL}}(T)$ is removed by extending the oscillating-wall concept: a time-dependent spanwise velocity profile of the form \eqref{eq:SL} is prescribed by superposing a time-varying spanwise body force onto the oscillating wall, so that $T$ and $\delta$ can be decoupled.
A series of direct numerical simulations is conducted to investigate the influence of each parameter on drag reduction.
Results indicate that conventional wall oscillations do not fully exploit the potential of spanwise wall forcing. Larger drag reductions and greater energy savings can be achieved by exploring regions of the $(T,\delta)$ parameter space that are inaccessible for the oscillating wall: $T_{opt}^+ \approx 100$ does not reflect any intrinsic property of wall turbulence, but rather arises from the constraint $\delta = \delta_{\text{\tiny \!\! SL}}(T)$.

\section{Methods}

Direct numerical simulations of turbulent channel flow with spanwise forcing are conducted to study how independent variations of the temporal and spatial scales of the forcing affect drag reduction.
Throughout this study $x, y, z$ and $u, v, w$ denote the streamwise, wall-normal, and spanwise directions and the corresponding velocity components. The operator $\aver{ \cdot}$ indicates averaging over the homogeneous $x,z$ directions and time; when present, a subscript indicates partial averaging, e.g. $\aver{ \cdot }_{xz}$ refers to a spatial mean along the homogeneous directions. 

The one-to-one correspondence \eqref{eq:deltaSL} between the Stokes layer thickness and the oscillation period is removed by imposing, at each time step during the simulations, that the spatially averaged profile $\aver{w}_{xz}(y,t)$ of the spanwise velocity component follows the extended Stokes layer (ESL) profile:
\begin{equation}
w_{\text{\tiny \!\! ESL}}(y,t) = A \exp \left( - \frac{y}{\delta} \right) \sin \left( \frac{2\pi}{T} t - \frac{y}{\delta} \right),
\label{eq:ESL}
\end{equation}
where $w_{\text{\tiny \!\! ESL}}$ is formally identical to the Stokes layer \eqref{eq:SL}, except that $\delta$ and $T$ are independent.
The spanwise profile \eqref{eq:ESL} is obtained by complementing the oscillating-wall boundary condition \eqref{eq:OW} with the following unsteady spanwise body force $f_z$:
\begin{equation}
\label{eq:force}
f_z(y,t) = \left (\frac{ \partial w_{\text{\tiny \!\! ESL}} }{ \partial t } - \nu \frac{ \partial^2 w_{\text{\tiny \!\! ESL}} }{ \partial y^2} \right ) + \left ( \frac{\partial \aver{vw}_{xz}}{ \partial y} \right ) ,
\end{equation}
which follows directly from the governing equation for the spanwise velocity averaged along the homogeneous directions. Its spatial and temporal derivatives are evaluated using the same discretization schemes of the DNS code. 

The study employs an in-house DNS solver for the incompressible Navier--Stokes equations, introduced by \cite{luchini-quadrio-2006} and written in the CPL Compiler and Programming Language \citep{luchini-2021}. The solver follows the pseudo-spectral approach, uses Fourier expansions in the homogeneous directions, and fourth-order compact explicit finite differences in the wall-normal direction. Temporal integration employs a third-order Runge--Kutta scheme for the nonlinear terms and a second-order Crank--Nicolson scheme for the viscous term.

The simulations are carried out at constant flow rate (CFR), with the bulk Reynolds number set at $Re_b = U_b h / \nu = 7000$, which corresponds to a friction Reynolds number of $Re_\tau = u_\tau h / \nu \approx 400 $ in the unforced case. 
Here, $ U_b $ is the bulk velocity, $ u_\tau = \sqrt{\tau_w / \rho} $ is the friction velocity defined in terms of the averaged wall-shear stress $ \tau_w $ and fluid density $\rho$, and $h$ is the channel half-height.
Note that, to compare with several existing oscillating-wall studies, a preliminary study was conducted at $Re_\tau = 200$. However, since the ESL is more effective than the conventional oscillating wall, relaminarisation occurred over a range of parameters, suggesting the adoption of a higher Reynolds number to obtain a clearer picture.

The computational domain has dimensions $(L_x, L_y, L_z) = (4 \pi h, 2h, 2 \pi h)$. It is discretised with $N_y = 400$ grid points in the wall-normal direction and $N_x = N_z = 512$ Fourier modes in the streamwise and spanwise directions. To remove aliasing errors, the Fourier modes are increased by a factor of $3/2$ when computing the nonlinear terms in physical space. With these extra modes, the streamwise and spanwise resolutions are $\Delta x^+ \approx 6.5$ and $\Delta z^+ \approx 3.3$, respectively.
The collocation points along the wall-normal direction follow a non-homogeneous hyperbolic tangent distribution, yielding a minimum wall-normal spacing near the wall of $ \Delta y^+_{\min} \approx 0.6$ and a maximum spacing at the channel centreline of $\Delta y^+_{\max} \approx 3.3$.
Simulations are conducted for a total time of $700 h / U_b$, with the first $100$ time units excluded from the computation of flow statistics to eliminate the initial transient.
The oscillation period is varied within the range $10 \leq T^+ \leq 200$, while the thickness varies in $2 \leq \delta^+ \leq 20$. The forcing amplitude is fixed at $A^+ = 12$. In total, the study includes $111$ simulations.

\begin{figure}
\centering
\includegraphics[width=0.8\textwidth]{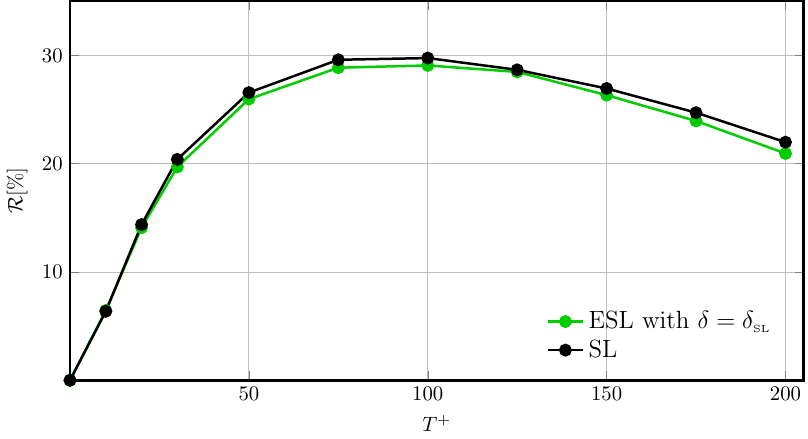}
\caption{Drag reduction $\DR$ versus oscillation period $T^+$ for the ESL forcing with $\delta=\delta_{\text{\tiny \!\! SL}}$ (green) and the oscillating wall (black), at $A^+=12$ and $Re_\tau=400$.}
\label{fig:validation}
\end{figure}

The primary quantity of interest is the drag reduction rate $\DR$, defined as
\begin{equation}
\DR = 100 \times \left( 1 - \frac{C_f}{C_{f,0}} \right),
\end{equation}
where $C_f = 2 \tau_w / (\rho U_b^2)$ is the skin-friction coefficient, and the subscript $0$ indicates the reference, uncontrolled case.

To validate the numerical approach, the drag reduction rate obtained with the standard oscillating wall is compared to that from the ESL \eqref{eq:ESL} when $\delta = \delta_{\text{\tiny \!\! SL}}(T)$, by performing two sets of simulations using identical numerical procedures. Figure~\ref{fig:validation} shows the drag reduction rates from the two sets and confirms their agreement.
The tendency of the ESL forcing to predict slightly lower drag reduction than an actual oscillating wall can be attributed to the known small residual differences \citep{quadrio-sibilla-2000, choi-xu-sung-2002, touber-leschziner-2012} between the laminar Stokes profile and the actual coherent spanwise profile, particularly at large $T$. In these cases, the laminar Stokes layer exhibits slightly smaller velocity amplitudes \citep[see e.g., figure~5 of][]{quadrio-sibilla-2000}.

\section{Drag reduction}

\begin{figure}
\centering
\includegraphics[width=0.9\textwidth]{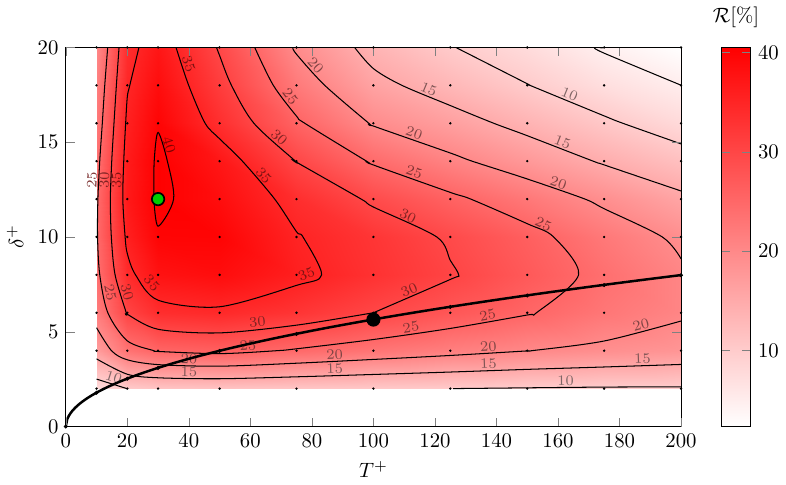}
\caption{Drag reduction map in the $(T,\delta)$ plane. The thick line indicates the $\delta=\delta_{\text{\tiny \!\! SL}}(T)$ constraint. The green/black dots identify the optimal point for ESL/SL.}
\label{fig:dr-map}
\end{figure}

Figure~\ref{fig:dr-map} shows the drag reduction map obtained with the ESL in the $(T,\delta)$ parameter plane. The tiny dots indicate the parameter pair for each simulation. The thick curve connects a subset of points where $\delta = \delta_{\text{\tiny \!\! SL}}(T)$; so far, wall oscillation studies could explore the parameter plane only along this line, hereafter referred to as the SL line.

In agreement with previous information for the oscillating wall, figure \ref{fig:dr-map} shows that, at $Re_\tau = 400$ and $A^+ = 12$, the maximum on the SL line (indicated by the large black dot) reaches $\DR \approx 30\%$ at $(T^+, \delta^+) \approx (100, 5.7)$. 
However, the ESL maximum drag reduction occurs far from the SL line, achieving $\DR \approx 41\%$ at $T^+ = 30$ and $\delta^+ = 12$, i.e., at lower $T^+$ and higher $\delta^+$. 
It is worth noting that, along the SL line, decreasing $T$ corresponds to decreasing $\delta$. 

On the drag reduction map, $\DR$ is largest over a relatively broad region, remaining above 30\% for $20 \le T^+ \le 50$ and $6 \le \delta^+ \le 20$, and significant drag reduction is also observed for larger periods when $\delta^+ \approx 8$. Values of $\delta^+$ exceeding the classical SL optimum at a given $T^+$ indicate that effective ESL configurations penetrate far enough from the wall to interact with turbulent structures in the buffer layer.
When $\delta$ is small, drag reduction is low regardless of $T$, as the forcing remains confined within the viscous sublayer, where turbulent activity is minimal. This confirms that effective control requires interaction with turbulence above the viscous sublayer; drag reduction is not achieved through direct manipulation of the streamwise velocity at the wall, where the wall shear is defined.
When the oscillation period is reduced below $T^+ \lesssim 20$, $\DR$ decreases regardless of $\delta$, indicating that the spanwise motion is too rapid to effectively interact with near-wall turbulence. This observation is consistent with the existence of a minimum period required for effective control.
Finally, as $T^+$ increases beyond $50$, the range of optimal $\delta^+$ narrows, for instance, $5 \lessapprox \delta^+ \lessapprox 12$ at $T^+ = 100$; the difference between the ESL and SL maxima gradually diminishes, and the best ESL configuration reduces to the SL.

\begin{figure}
\centering
\includegraphics[width=\textwidth]{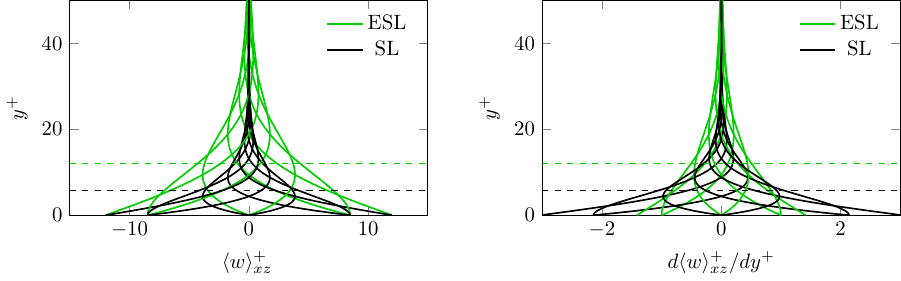}
\caption{Comparison between the optimal ESL (green) and SL (black) spanwise velocity profiles. Left: wall-normal distribution of the spanwise velocity component; right: wall-normal distribution of the spanwise shear. The dashed lines mark the location of $\delta_{opt}$ for the two cases.}
\label{fig:ESLvsSL}
\end{figure}
Figure~\ref{fig:ESLvsSL} illustrates the SL and ESL profiles in their respective optimal configurations, along with the corresponding wall-normal distributions of spanwise shear. The two profiles therefore correspond to different combinations of $T$ and $\delta$. The spanwise motion induced by the best-performing ESL remains significant up to larger wall distances, consistent with the larger $\delta_{opt}$. Since the maximum wall velocity is the same, the increased thickness reduces the spanwise shear near the wall for the ESL, but produces higher spanwise shear in the buffer layer.

\begin{figure}
\centering
\includegraphics[width=0.9\textwidth]{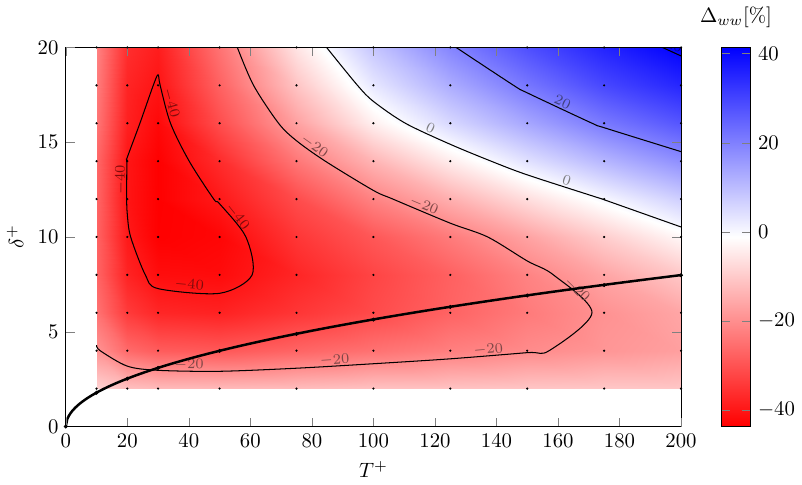}
\caption{Map in the ($T,\delta)$ plane of the integral change $\Delta_{ww}$ of the variance of the spanwise fluctuations with respect to the reference case. The thick line indicates the $\delta=\delta_{\text{\tiny \!\! SL}}(T)$ constraint. Note the inverted color scale.} 
\label{fig:ww-map}
\end{figure}

The amount of drag reduction is governed by the ability of the forcing to generate a spanwise velocity profile with minimal fluctuations around the spatially averaged profile $\langle w \rangle_{xz}(y,t)$. This property, already discussed in wall-oscillation studies \citep[e.g.,][]{quadrio-ricco-2011, touber-leschziner-2012}, becomes more evident here once the constraint of injecting spanwise forcing only through the wall movement is removed.
The deviation of the local spanwise profile from $\aver{w}_{xz}$ is quantified by $\Delta_{ww}$, defined as the integral across the channel of the variance of the spanwise velocity fluctuations around the spatial mean, minus the variance $\langle w^2 \rangle_0$ of the spanwise velocity fluctuations in the reference flow, and expressed as a percentage of the latter:
\begin{equation}
\Delta_{ww} = 100 \times \frac{ \int_0^{2h} \left[ \aver{ \left ( w - \aver{w}_{xz} \right)^2} - \aver{w^2}_0 \right] dy}{\int_0^{2h} \aver{w^2}_0 \, dy} .
\end{equation}

Figure \ref{fig:ww-map} plots the distribution of $\Delta_{ww}$ in the parameter space. The contour plot resembles the drag reduction map in figure~\ref{fig:dr-map}: the region with strong suppression of spanwise turbulent fluctuations, where the variance is up to $40\%$ lower than in the unperturbed flow, corresponds to large drag reduction, whereas the area with $\Delta_{ww} > 0$, indicating increased spanwise fluctuations, coincides with vanishing drag reduction.

\section{Power budget}
\label{sec:power}

\begin{figure}
\centering
\includegraphics[width=0.9\textwidth]{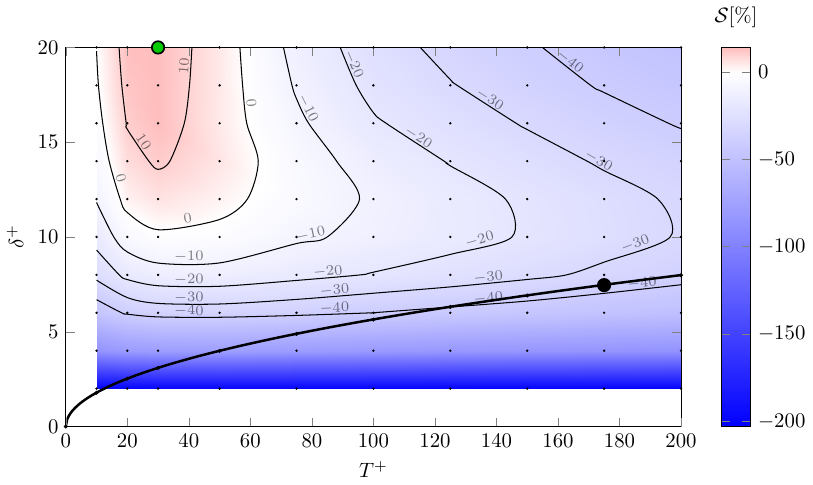}
\caption{Net power saving map in the $(T,\delta)$ plane. The thick line indicates the $\delta=\delta_{\text{\tiny \!\! SL}}(T)$ constraint. The green/black dots identify the optimal point for ESL/SL.}
\label{fig:s-map}
\end{figure}

The qualitative observation that the most effective spanwise forcing lies far from the SL line is further supported by considering the net energy saving potential of the ESL, evaluated under the usual assumption of ideal actuation \citep{baron-quadrio-1996}.

The control power $\mathcal{P}_c$, i.e., the energy per unit time required to generate the ESL spanwise velocity profile, has two contributions: the power $P_w$ needed to move the wall against the viscous fluid, and the power $P_f$ associated with the spanwise body force. In the conventional SL, only $P_w$ appears, which is known \citep{choi-xu-sung-2002} to be well represented via the analytical expression of the laminar Stokes layer.

Powers are expressed as percentage of the pumping power $P_0$ of the uncontrolled flow, given by $P_0 = 2 L_x L_z U_b \aver{\tau_x}$, as:
\begin{equation}
  \mathcal{P}_c = P_w + P_f = 100 \times \frac{L_x L_z}{P_0} \left( \aver{ 2 w_w \tau_z } + \rho \int_{0}^{2h} \aver{ f_z   w}  dy \right)
\label{eq:power}
\end{equation}
where $\tau_z$ is the spanwise component of the instantaneous wall shear stress. The net energy saving rate $\mathcal{S}$ is computed as $ \mathcal{S} = \DR - \mathcal{P}_c $.

Consistent with previous results, at $Re_\tau = 400$ and $A^+ = 12$, the oscillating wall produces $\mathcal{S} < 0$ for all $T^+$: along the black line in figure~\ref{fig:s-map}, the best outcome is $\mathcal{S} = -35\%$, achieved at $T^+ = 175$ and $\delta^+ = 7.46$. In contrast, the ESL yields clearly positive net savings, even at this high forcing amplitude: $\mathcal{S} \approx 16\%$ is achieved for $T^+ = 30$ and the largest $\delta^+$ considered.

\begin{figure}
\centering
\includegraphics[width=\textwidth]{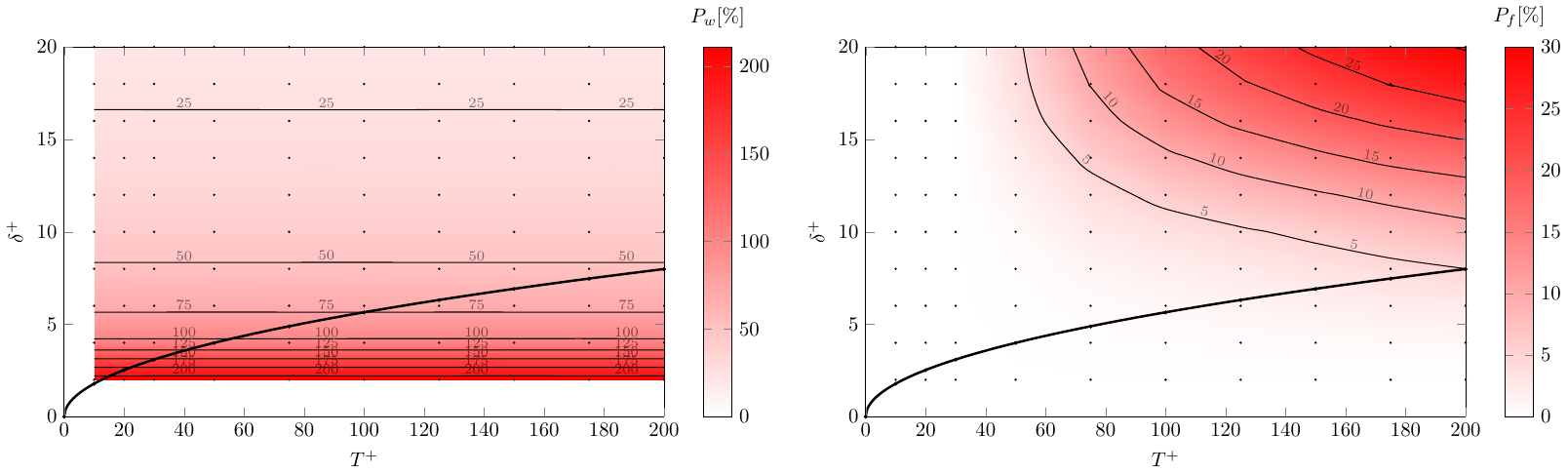}
\caption{Control power map in the $(T,\delta)$ plane: $P_w$ (left) and $P_f$ (right) expressed as percentages of the pumping power $P_0$. The thick line indicates the $\delta=\delta_{\text{\tiny \!\! SL}}(T)$ constraint. }
\label{fig:powers}
\end{figure}

The two contributions to control power are shown separately in figure \ref{fig:powers}.
For fixed forcing amplitude, $P_w$ (left panel) depends only on $\delta$ and decreases as the thickness increases, since a larger $\delta$ corresponds to a smaller spanwise wall shear. Therefore, the ESL ability to operate effectively at large $\delta$ is key to improving the net energy balance. On the right panel, $P_f$ is shown to remain small compared to $P_0$, except for the top-right area.

Comparing the two contributions to $P_c$ shows that, over most of the parameter space, $P_f < P_w$, often by a substantial margin. Only when both $T$ and $\delta$ are large, $P_f$ becomes the dominant contribution. Moreover, although the instantaneous power associated with the forcing terms in \eqref{eq:force} containing the ESL profile can be large, the periodicity of the ESL implies that their time average vanishes. As a result, the last term in \eqref{eq:force}, which accounts for nonlinear turbulence interactions, provides the only net contribution to $P_f$.
At the SL local optimum, $P_w$ is approximately $56\%$ of the pumping power; at the ESL optimum, this fraction drops to $19\%$, while $P_f$ contributes $7\%$.

\section{Concluding discussion}
\label{sec:fs_conclusion}

A DNS study of turbulent channel flows with spanwise forcing for drag reduction has been conducted to disentangle the effects of the forcing period $T$ and the penetration depth $\delta$ of the transversal Stokes layer (SL).
 
In the conventional oscillating wall setup, the two parameters are not independent, and the SL thickness is determined by $T$ via $\delta = \delta_{\text{\tiny \!\! SL}}(T)$. Here, instead, a time-varying spanwise volume force is applied to the momentum equation, in addition to wall motion, to create an extended Stokes layer (ESL) in which $\delta$ and $T$ can be varied independently. The bulk Reynolds number is fixed at $Re_b = 7000$ ($Re_{\tau,0} = 400$). A total of 111 simulations are performed, spanning $10 \le T^+ \le 200$ and $2 \le \delta^+ \le 20$, with the wall oscillation amplitude held constant at $A^+ = 12$.

The ESL achieves significantly higher drag reduction than the SL, with the maximum increasing from $30\%$ to $41\%$. The ESL optimum occurs at markedly different parameter values: $(T^+_{opt}, \delta^+_{opt}) \approx (30, 12)$ for the ESL, compared to $(T^+_{opt}, \delta^+_{opt}) \approx (100, 5.7)$ for the SL. Moreover, the ESL yields a net positive energy saving, unlike the SL, whose net savings are always negative at this forcing intensity.

\begin{figure}
\centering
\includegraphics[width=\textwidth]{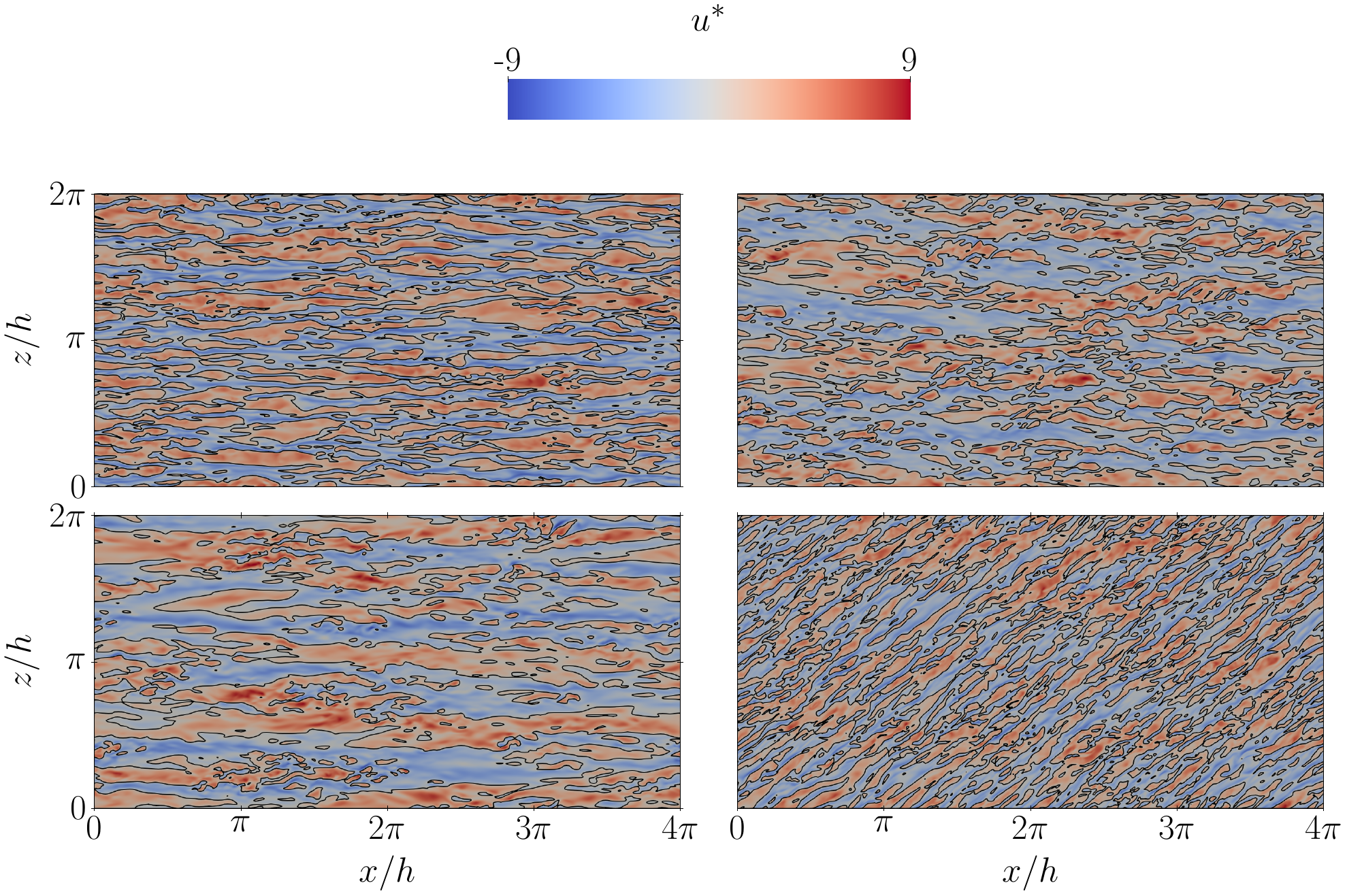}
\caption{Snapshots of the field of streamwise velocity fluctuations in a wall-parallel plane at $y^*=14.5$. Top left: reference flow; top right: optimal SL; bottom left: optimal ESL; bottom right: suboptimal ESL. All controlled cases are at the same (zero) phase.}
\label{fig:fields}
\end{figure}

Looking at instantaneous flow fields (figure \ref{fig:fields}, which plots wall-parallel slices at the wall-distance $y^*=14.5$ where the fluctuations without forcing are largest) suggests that the ESL does not radically alter the drag reduction process. Quantitative differences between SL and ESL therefore descend from the different relative weight of $T$ and $\delta$, and from their consequent effects on the canonical wall turbulence cycle. The reference case (top left) shows the expected streaky structure of alternating low- and high-velocity streaks, which are significantly reoriented in the suboptimal ELS case ($T^+ = 200, \delta^+ = 12$, bottom right), as the flow in the buffer layer directly couples to the slow spanwise motion \citep{touber-leschziner-2012}. The optimal SL $(T^+ = 100, \delta^+ = 5.6)$ and ESL $(T^+ = 30, \delta^+ = 12)$ cases are quite similar to each other: the large drag reduction translates into larger spatial scales, and there is no evidence of streak reorientation, as already observed in previous studies \citep{quadrio-ricco-viotti-2009} for spanwise wall forcing near its optimum configuration.

\begin{figure}
\centering
\includegraphics[width=0.8\textwidth]{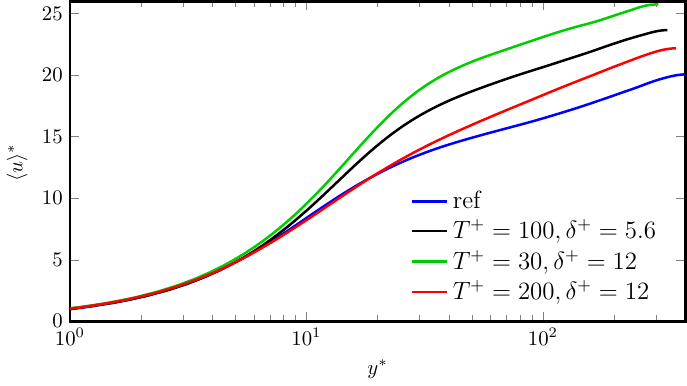}
\includegraphics[width=0.32\textwidth]{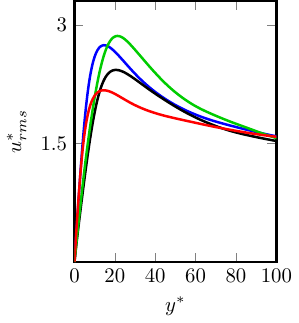}
\includegraphics[width=0.32\textwidth]{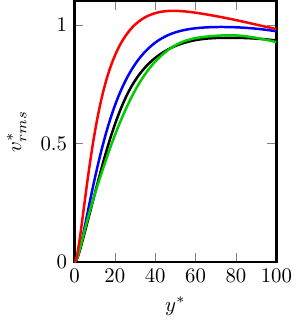}
\includegraphics[width=0.32\textwidth]{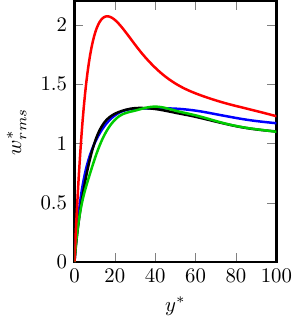}
\caption{Longitudinal mean velocity profiles (top) and r.m.s. value of velocity fluctuations (bottom): optimal SL and ESL are compared to the reference case and to a suboptimal ESL case at large $T$.}
\label{fig:stats}
\end{figure}

Figure \ref{fig:stats} compares the mean velocity profiles and the root mean squared values of the velocity fluctuations. The mean profiles present the expected upward shift in the logarithmic region, as a consequence of the reduced drag, the only difference between ESL and SL being the large vertical shift of the former, as a consequence of the larger drag reduction. The suboptimal ESL displays a vertical displacement but also an increase of the slope of the logarithmic region. 
Figure \ref{fig:stats} also shows profiles of r.m.s. of velocity fluctuations. The $u$ profiles are in agreement with the instantaneous fields shown in figure \ref{fig:fields}, are quite similar between SL and ESL, and show a peak that is shifted outwards. The same is true for the other two components, which perfectly overlap when plotted in actual viscous units. 
The combinations of $T$ and $\delta$ that outperform the SL parameters do not seem to involve novel drag reduction physics on top of the one at work with the standard SL. Hence, the reason why removing the constraint \eqref{eq:deltaSL} leads to a more effective forcing remains to be fully understood. By better understanding the physics involved, one could then devise a rational procedure to convert the ESL profile into an optimal forcing.

In conclusion, this study shows that the physical interpretation of the well-known optimum period $T^+_{opt} = 100$ for the oscillating wall should be reconsidered: this value appears to stem from the constraint $\delta = \delta_{\text{\tiny \!\! SL}}(T)$ imposed by the oscillating wall, and thus from a specific implementation of spanwise forcing, rather than from intrinsic flow properties. 
Indirectly, these results suggest searching for alternative actuation strategies to generate near-wall spanwise motion. Such strategies should be able to remove the constraint $\delta = \delta_{\text{\tiny \!\! SL}}(T)$, thereby fully unlocking the drag-reduction potential of spanwise forcing. Promising (active and passive) strategies already exist, and include among others plasma actuators \citep{jukes-choi-2012, benard-etal-2021,thomas-etal-2019}, electromagnetic tiles \citep{du-karniadakis-2000}, electroactive polymers combined with electromagnetic actuators \citep{gouder-potter-morrison-2013, gatti-etal-2015}, alternating slip/no-slip stripes \citep{fuaad-arulprakash-2019}, surface dimples \citep{gattere-chiarini-quadrio-2022}, sinusoidal riblets \citep{peet-sagaut-charron-2008}. However, their design could be improved by clearly understanding the target physical process that must be activated in the flow to achieve drag reduction. 
Regardless of a specific actuation technology, the present study sets up a new framework to examine the physical effects leading to drag reduction, and will be useful to design actuators intended to implement them.


\section*{Acknowledgments} 
Preliminary versions of this work have been presented by M.Q. at the XVIII Euromech Turbulence Conference in Valencia, September 2023; by F.G. at the European Drag Reduction and Flow Control Meeting in Torino, September 2024, and at the ITI XI in Bertinoro, July 2025. This material is based upon work supported by the Air Force Office of Scientific Research (M.Q., award number FA8655-25-1-7006).

\section*{Declaration of Interests} 
The authors report no conflict of interest.

\bibliographystyle{jfm}
\bibliography{../Wallturb}

\end{document}